\documentstyle[aps,floats,graphicx,amssymb]{revtex}

\begin{document}
\tightenlines
\def\Vec#1{\mbox{\boldmath $#1$}}
\twocolumn \noindent {\Large\bf Transition to superfluid turbulence governed by an
intrinsic parameter}
\begin{flushleft}
  \bf
  A.P.~Finne$^*$, T.~Araki$^{\S}$, R.~Blaauwgeers$^{*,\dagger}$,
V.B.~Eltsov$^{*,\ddagger}$,
  N.B.~Kopnin$^{*,\|}$, M.~Krusius$^*$, L.~Skrbek$^1$, M.~Tsubota$^{\S}$,
\& G.E.~Volovik$^{*,\|}$
\end{flushleft}
\noindent
{\it
  $^*$ Low Temperature Laboratory, Helsinki University of Technology,
  P.O.Box 2200, FIN-02015 HUT, Finland.\\
  $^{\S}$ Department of Physics, Osaka City University, Sumiyoshi-Ku,
  Osaka 558-8585, Japan.\\
  $^\dagger$ Kamerlingh Onnes Laboratory, Leiden University, P.O.Box
  9504, 2300 RA Leiden, The Netherlands.\\
  $^\ddagger$ Kapitza Institute for Physical Problems, Kosygina 2, 119334
  Moscow,  Russia.\\
  $^{\|}$ Landau Institute for Theoretical Physics, Kosygina 2, 119334
  Moscow, Russia.\\
  $^1$ Joint Low Temperature Laboratory, Institute of Physics ASCR and
  Charles University, Prague, Czech Republic.\\
}

{\bf Hydrodynamic flow in both classical and quantum fluids can be either laminar or
turbulent. To describe the latter, vortices in turbulent flow are modelled with
stable vortex filaments. While this is an idealization in classical fluids, vortices
are real topologically stable quantized objects in superfluids. Thus superfluid
turbulence \cite{VinenNiemela} is thought to hold the key to new understanding on
turbulence in general. The fermion superfluid $^3$He offers further possibilities
owing to a large variation in its hydrodynamic characteristics over the
experimentally accessible temperatures. While studying the hydrodynamics of the B
phase of superfluid $^3$He, we discovered a sharp transition at 0.60 T$_{\rm c}$
between two regimes, with regular behaviour at high-temperatures and turbulence at
low-temperatures. Unlike in classical fluids, this transition is insensitive to
velocity and occurs at a temperature where the dissipative vortex damping drops
below a critical limit. This discovery resolves the conflict between existing high-
and low-temperature measurements in $^3$He-B: At high temperatures in rotating flow
a vortex loop injected into superflow has been observed to expand monotonically to a
single rectilinear vortex line \cite{Single-vortex,Ruutu,KH-PRL}, while at very low
temperatures a tangled network of quantized vortex lines can be generated in a
quiescent bath with a vibrating wire \cite{Fisher}. The solution of this conflict
reveals a new intrinsic criterion for the existence of superfluid turbulence. }

\vspace{2mm}

In conventional liquids the vorticity
$\Vec{\omega}=\Vec{\nabla}\times {\bf v}$ obeys the Navier-Stokes
equation
\[ 
\frac{\partial \Vec{\omega}}{\partial t}= \Vec{\nabla} \times
[{\bf v}\times
\Vec{\omega}]+  \nu\Vec{\nabla}^2 \Vec{\omega}\,. 
\] 
The interplay of the two terms on the rhs, the inertial first term and the viscous
second term, governs the transition to turbulence \cite{McComb}. It is determined by
the external conditions through the Reynolds number, $Re=RU/\nu$, formed by the
characteristic size $R$ of the system, the flow velocity $U$, and the kinematic
viscosity $\nu$. At large $Re\gg 1$, the effect of the inertial term is dominating,
and laminar flow becomes increasingly unstable. If vorticity is released into a
meta-stable laminar state, a sudden transition to a chaotic flow of eddies occurs.
The evolution of the turbulent flow is described by the Kolmogorov energy cascade:
the kinetic energy of the flow is transferred to ever smaller length scales, with
large vortex loops decaying into smaller loops, until a scale is reached where the
energy is dissipated by viscosity.

In superfluids turbulence acquires new features: A superfluid can be described as
consisting of two inter-penetrating fractions, the frictionless (superfluid) and
viscous (normal) components. If both fractions are moving, as is the case {\em eg.}
behind a pulled grid in measurements on superfluid turbulence in $^4$He
\cite{VinenNiemela}, the turbulent state bears more resemblance to that of viscous
liquids. A new class of superfluid turbulence becomes possible when the normal
component is so viscous that it is essentially immobile and only the superfluid
component is moving with respect to the boundaries. This is the usual case in the
flow of $^3$He, which is considered here.

\begin{figure}[t!!]
\begin{center}
\leavevmode
\includegraphics[width=\linewidth]{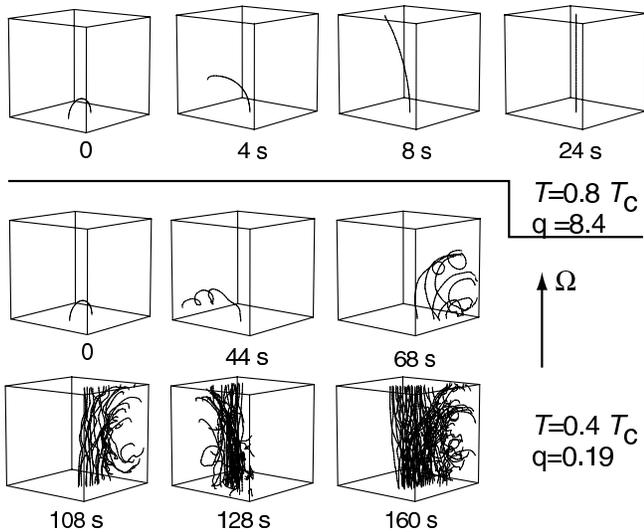}
\caption{Summary of events: A vortex half ring is injected into {\em vortex-free}
superflow, generated with rotation around the vertical axis ($\Omega = 0.21\,$rad/s
or ${\rm Re}_{\rm s} \approx 30$). The figures represent snapshots from simulation
calculations in the rotating frame, which are consistent with the experimental
observations. The only temperature dependence is contained in the mutual-friction
parameters $\alpha$ and $\alpha'$ which we take from Ref.~\protect\cite{Bevan}. {\it
(Top)} At high temperatures the loop grows monotonically into a single rectilinear
vortex line, pulled by the Magnus force from the superflow. The motion is highly
damped since the Magnus force is balanced by the mutual friction force. {\it
(Bottom)} At low temperatures the loop moves with the circulating superflow: (44\,s)
The vortex is unstable with respect to loop formation and develops through
reconnections into a tangle at the injection site. (68\,s) The Magnus force from the
vortex-free superflow extends the tangle along the vertical axis. The measured
flight time $\tau_{F}$ for the tangle to extend across a vertical distance $d$ is
controlled by mutual-friction damping: $\tau_{F} \approx d/(\Omega R \alpha)$. (108
--- 160\,s) The final step is the reconnection of the tangle to rectilinear lines.
The growth in the relative polarization is experimentally observed as a relaxation
in the NMR absorption and the total number of vortex lines in the final stable state
is determined from the NMR line shape (Fig.\,\protect\ref{fillingratio}). }
\label{Simulation}
\end{center}
\end{figure}

The vorticity of the superfluid component is quantized in terms of
the circulation quantum $\kappa =2\pi \hbar /M$ where $M$ is the
mass of a superfluid particle. The flow of the superfluid
component can be characterized with the `superfluid Reynolds
number' $Re_{\rm s}=RU/\kappa$. The Feynman criterion $Re_{\rm s}
\sim 1$ gives the velocity at which it becomes energetically
favorable to form a quantized vortex. If a large nucleation
barrier exists, vortices are not created and the superfluid
remains vortex-free, in our measurements here up to $Re_{\rm s}
\sim 200$. Quantized vortices can then be injected, either by some
extrinsic means or by increasing the velocity so that the
nucleation barrier is overcome and spontaneous vortex formation
occurs. The limit of large $Re_{\rm s}\gg 1$ is equivalent to a
vanishing Planck's constant, $\kappa \propto \hbar \rightarrow 0$,
so that the vorticity becomes a continuous variable, like in the
classical case. However, the fluid still remains unconventional
because of its two-fluid nature: In addition to the superflow,
there is the normal component which is at rest in the container
frame, ${\bf v}_{\rm n}=0$. Interactions between the superfluid
vorticity and the normal component give rise to a mutual friction
force on a unit volume of the superfluid,
\[
{\bf f}_{\rm mf} =-\alpha ^\prime \rho _{\rm s} [{\bf v}_{\rm
s}\times \Vec{\omega}_{\rm s}]+\alpha \rho_{\rm
s}[\hat{\Vec{\omega}}_{\rm s}\times [\Vec{\omega}_{\rm s} \times
{\bf v}_s]]\,,
\]
which is described by the dimensionless temperature-dependent parameters $\alpha$
and $\alpha'$. These correspond to the usual mutual friction parameters, if we are
dealing with one vortex or locally polarized vorticity, or they may be renormalized
in the general case. Here ${\bf v}_{\rm s}$ and $\Vec{\omega}_{\rm s}=\Vec{\nabla}
\times {\bf v}_{\rm s}$ are the superfluid velocity and vorticity, and
$\hat{\Vec{\omega}}_{\rm s} $ is the unit vector in the direction of
$\Vec{\omega}_{\rm s}$. Inserting the mutual-friction force in the Euler equation
one obtains the hydrodynamic equation for $Re_{\rm s} \gg 1$ \cite{Sonin}:
\begin{equation}
\frac{\partial \Vec{\omega}_{\rm s}}{\partial t} = (1-\alpha')\Vec{\nabla }\times
[{\bf v}_{\rm s} \times \Vec{\omega}_{\rm s}]+ \alpha \Vec{\nabla }\times
[\hat{\Vec{\omega}}_{\rm s} \times(\Vec{\omega}_{\rm s} \times{\bf v}_{\rm s})] \ .
\label{SuperfluidHydrodynamics}
\end{equation}
The energy dissipation is determined by the mutual friction damping $\alpha$ in the
second term, while the reactive mutual-friction $\alpha'$ renormalizes the inertial
term of conventional hydrodynamics. Like in a classical fluid, the inertial term
drives the instability by increasing the number of vortex loops, whereas the
dissipative term acts to stabilize laminar flow by decreasing the number of loops.
The fundamental difference from conventional hydrodynamics is that these two
competing terms now have a similar dependence on the velocity and its gradients.
Their relative importance is determined by the intrinsic dimensionless parameter of
the superfluid, $q=\alpha/(1-\alpha')$, in contrast to classical liquids where it is
governed via the Reynolds number by the extrinsic quantities $R$ and $U$. We thus
expect laminar flow to become unstable when $q$ drops below a critical value
$q_c\sim 1$. In the case of a single vortex line the parameter $q$ also marks the
crossover from propagating Kelvin waves $(q \lesssim 1)$ to the over-damped regime
$(q \gtrsim 1)$ \cite{Barenghi}. In Fermi superfluids and superconductors with
finite energy gap one has $q\approx (\omega_0\tau)^{-1}$, where $\omega_0$ is the
spacing between the bound states of quasiparticles in the vortex core and
$\tau^{-1}$ is their lifetime broadening, owing to scattering from the normal
component \cite{KopninBook}. In $^3$He-B, $q$ approaches infinity at $T_{\rm c}$ and
drops monotonically to zero, when cooled down to $T\rightarrow 0$
\cite{KopninBook,Bevan}. Such an evolution of $q$ with temperature would in
classical fluids correspond to scanning the Reynolds number from 0 to $\infty$: We
can thus study the emergence of turbulence as a function of temperature when $q$
falls below unity.

\begin{figure*}[t!!]
\begin{center}
\leavevmode
\includegraphics[width=1\linewidth]{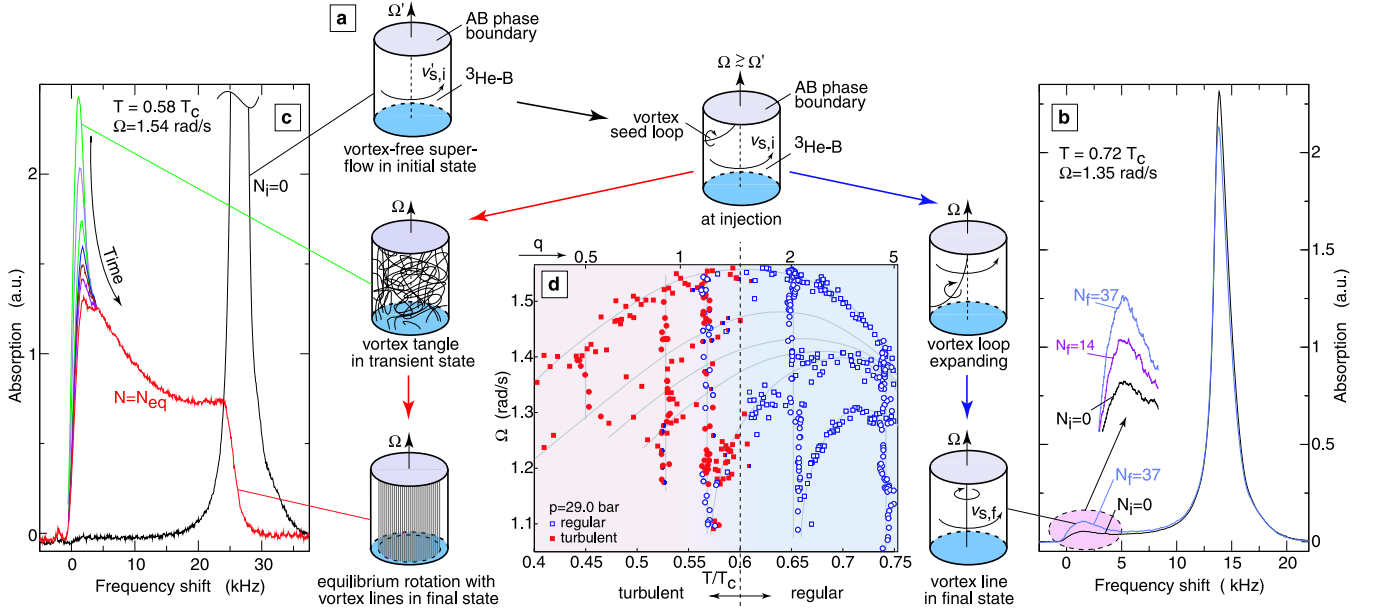}
\vspace{3mm} \caption{Principle of measurement and central result: {\bf a}, The
initial state is vortex-free superflow in rotation, with the normal component
stationary and the superfluid component moving in the rotating frame. This state is
identified by a large sharp peak in the NMR absorption spectrum, which is caused by
the vortex-free superflow and is shifted far from the Larmor value (line shape
denoted as $N_{\rm i}=0$ in {\bf b} and {\bf c}). When a few $(\Delta N)$ vortex
loops are injected, the number of vortex lines in the final steady state $(N_{\rm
f})$ is found to fall in one of two categories: {\bf b}, $N_{\rm f} = \Delta N$,
indicated by a small shift of the NMR absorption from the superflow peak close to
the Larmor value. The shifted absorption is proportional to $N_{\rm f}$, as shown in
the insert by the build up of absorption at small frequency shift. This process
corresponds to regular loop expansion. {\bf c}, $\Delta N \ll N_{\rm f} \lesssim
N_{\rm eq}$, indicated by a total removal of the superflow peak {\it (red)}. Here
$N_{\rm eq} \approx \pi R^2 \, 2\Omega/\kappa \sim 10^3$ is the equilibrium number
of vortex lines in rotation. This process corresponds to turbulent loop expansion.
The relaxation of the line shape through intermediate stages ({\bf c}) represents a
direct signal from the turbulent epoch: In the final state with a large number of
rectilinear vortex lines $(N_{\rm f} \approx N_{\rm eq})$, the distribution of
absorption is relatively flat. In contrast, in the turbulent state with unpolarized
${\bf v}_{\rm s}$ field, much of the absorption is piled in a narrow peak at the
Larmor edge. The intermediate curves between these two limiting line shapes
represent the transient relaxation when the turbulent vorticity with $\omega_{\rm s}
\gg \omega_{\rm eq}$ decays to a polarized cluster of rectilinear lines with
$\omega_{\rm eq} = 2\Omega$. {\bf d}, Phase diagramme of measured events: Each data
point monitors the outcome from an injection event which is initiated using the
shear-flow instability of the phase boundary between $^3$He-A and $^3$He-B. Vortex
loop expansion separates into regular events {\it (blue)} and turbulent events {\it
(red)} as a function of rotation $\Omega$ and temperature $T$, with an abrupt
transition at $0.60\,T_{\rm c}$ independently of $\Omega$. Only within a very narrow
interval $\Delta T \sim 0.05\,T_{\rm c}$ around $0.60\,T_{\rm c}$ can intermediate
values of $N_{\rm f}$ be measured. By scanning the temperature (curved lines) or the
current in the magnet for stabilizing the AB phase boundary (vertical lines), the
critical velocity $v_{\rm c} \approx \Omega R$ of the interface instability can be
varied along well-defined continuous trajectories. } \label{fillingratio}
\end{center}
\end{figure*}

{\it Measurements:}---In rotating $^3$He-B high-velocity vortex-free superflow can
be achieved, into which vortex seed loops can be injected (Fig.\,\ref{Simulation}).
We record as a function of temperature $T$ and rotation $\Omega$ the number of
rectilinear vortex lines after injection $(N_{\rm f})$, when all transients have
died out. Non-invasive NMR measurement is used to determine the number of vortex
lines in the final state. The NMR absorption spectrum of $^3$He-B maps the
order-parameter texture, which is strongly affected by the orienting effect from the
${\bf v}_{\rm s}$ field. It allows us to determine $N_{\rm f}$ with a resolution
better than 10 vortex lines in the temperature range of Fig.~\ref{fillingratio}. In
Fig.\,\ref{fillingratio}d we list the result after each injection event in the
$(\Omega,T)$ plane, classified as a regular high-temperature {\it (blue)} or
turbulent low-temperature {\it (red)} process. Here the injection method is the
shear-flow instability of the phase boundary between the $^3$He-A and $^3$He-B
phases, which produces a small random number ($\Delta N \sim 10$) of B-phase vortex
loops whose one end sticks out of the AB interface while the other end is on the
cylindrical sample boundary \cite{KH-PRL}. At high temperatures an injected loop
expands and becomes a single rectilinear vortex line. At low temperatures the loop
evolves into tangled vorticity which finally fills the whole sample volume with
rectilinear lines. The central plot Fig.\ \ref{fillingratio}d leads to a surprising
conclusion: Only a narrow cross-over region at $0.60\,T_{\rm c}$ separates the two
processes, with little dependence on the maximum superflow velocity $U=\Omega R$ of
the initial vortex-free state. At the transition temperature $0.60\,T_{\rm c}$, the
experimental value of $q=\alpha/(1-\alpha') \approx 1.3$ \cite{Bevan}.

Vortices can also be injected with other methods. In the absence of the AB interface
vortices are created (i) at a rough spot on the cylindrical sample boundary
\cite{Single-vortex}, (ii) under neutron irradiation \cite{Ruutu}, or (iii) by a
sudden leak of vortex loops through the small orifice which connects the sample to
the rest of the liquid $^3$He volume \cite{Leak}. These three processes have
different critical velocities and temperature dependences, but in all cases the
measurements are consistent with a crossover at $0.60\,T_{\rm c}$ from a low-yield
to a high-yield process with decreasing temperature. Depending on sample geometry,
below $0.50\,T_{\rm c}$ the vortex leak through the orifice may occur at 0.5\,rad/s
or less. Even at these low velocities $(Re_{\rm s} \lesssim 70)$ vortex loop
expansion is found to be turbulent.

The long cylindrical sample is monitored with two signal coils at different
locations, connected to independently operated NMR spectrometers. In addition to the
number of vortex lines in the final stable state, we can obtain information on the
transient state: (i) on the axial expansion of the vorticity and (ii) below
$0.60\,T_{\rm c}$ on the decay of the tangle into a polarized array of rectilinear
lines. The former process, the expansion of the vorticity along the $\hat
{\Vec{\Omega}}$ axis between the two signal coils, is controlled by the
mutual-friction damping $\alpha$: The extracted flight time for the motion in the
axial direction reproduces $\alpha(T)$ from Ref.~\cite{Bevan} and is continuous
across the transition at $0.60\,T_{\rm c}$. Thus the axial expansion is indifferent
to whether the vorticity moves as noninteracting loops or as a network of loops.

Below $0.60\,T_{\rm c}$, the arrival of the tangled vorticity within a pick-up coil
is signalled by a massive abrupt shift of the absorption close to the Larmor edge,
where a new sharp absorption peak is formed (Fig.~\ref{fillingratio}c). This line
shape is similar to that measured in the non-rotating state $(\Omega = 0)$. Here the
orientational effects from the superflow fields trapped around the vortex cores are
averaged out, owing to the randomly distributed vortex tangle with density
$\omega_{\rm s} \gg \omega_{\rm eq}$ and the removal of essentially all large-scale
vortex-free superflow. The sharp peak then decays with a relaxation time $\sim
30\,$s towards the broad final stable state spectrum of a polarized vortex array
with rectilinear lines and $\omega_{\rm s}= \omega_{\rm eq} = 2 \Omega$. This
restructuring of the NMR line shape is thus a direct signal from the transient epoch
with a vortex tangle.

{\em Numerical simulation}---As seen in Fig.~\ref{Simulation}, the measured features
are consistent with our simulation calculations. We use the vortex filament model
\cite{Schwarz1} in the rotating frame \cite{Araki}. A vortex is represented in
parametric form by ${\bf s}={\bf s}(\xi,t)$, where ${\bf s}$ refers to a point on
the filament and $\xi$ is the arc length along it. The spatial and time evolutions
are integrated rigorously using the Biot-Savart law. The vortex velocity $\dot{\bf
s}$ is calculated from the dynamic equation $\dot{\bf s}={\bf v}_{\rm sl} +\alpha
^\prime {\bf s}^\prime \times [{\bf s}^\prime \times {\bf v}_{\rm sl}]- \alpha [{\bf
s}^\prime \times {\bf v}_{\rm sl}]$ (recall that ${\bf v}_{\rm n}=0$), where the
local superfluid velocity ${\bf v}_{\rm sl}$ includes all contributions to the
superflow at ${\bf s}(\xi,t)$. As boundary conditions we use smooth solid-walls with
image vortices.

The initial state is a vortex half ring (diameter 3.5 mm) in the centre of a cubic
sample container (10\,mm wide). At high temperatures $T/T_{\rm c} = 0.8$ the high
damping $\alpha$ causes the seed loop to expand, mostly by motion in the radial and
axial directions, into a single rectilinear vortex line in the centre of the sample.
At low temperatures $T/T_{\rm c} = 0.4$ the seed loop travels azimuthally and
undergoes the Kelvin-wave instability, when the superflow is oriented along the
vortex \cite{Barenghi}. Here $q<1$ so that Kelvin wave oscillations are only lightly
damped and loop formation starts via reconnections. Thus a vortex network develops
already at the injection site. The vortex-free superflow above the tangle stretches
the loops with right orientation, while others shrink, and thus the tangle extends
and travels in the axial direction. Through reconnections the properly oriented
loops are gradually joined to lines which form a bundle almost parallel to the
rotation axis $\hat {\Vec{\Omega}}$. The bundle precesses initially around the $\hat
{\Vec{\Omega}}$ axis, but presumably more reconnections of remaining loops will
continue adding lines to the bundle and thereby reducing the superflow around the
bundle, until the equilibrium number of rectilinear vortex lines is reached. The
complete calculation is so far too time consuming.

{\em Discussion:}---Both the results from the numerical simulations in
Fig.~\ref{Simulation} and from the measurements in Fig.~\ref{fillingratio} support
the conclusion from Eq.~(\ref{SuperfluidHydrodynamics}) that the transition to
turbulence is determined by the parameter $q=\alpha/(1-\alpha')$: There exists a
critical value $q_{\rm c} \approx 1$ which separates the mutual-friction-dominated
regime at $q>q_{\rm c}$ from the inertia-dominated turbulent regime at $q<q_{\rm
c}$. With $q>q_{\rm c}$, the vortex-loop expansion is a regular stable process even
in high-velocity superflow with $Re_{\rm s} \gg 1$, as has been verified in many
previous measurements. With $q<q_{\rm c}$, a multiplication process is switched on,
the vortex lines create loops, become entangled, and reconnect forming new loops.
Could $q_{\rm c}$ be a fundamental universal number which, when $Re_{\rm s} \gg 1$,
places the upper bound to superfluid turbulence?

Why has such a transition, driven by an intrinsic parameter, not been observed
previously in superfluid $^4$He? A number of reasons can be listed: (i) In $^3$He
the viscosity of the normal fraction is four orders of magnitude higher. Thus there
is no ambiguity about the state of the normal component, unlike in the case of
$^4$He. (ii) The vortex-core radius in $^3$He-B exceeds by two orders of magnitude
that of $^4$He, where it is only of atomic size. Therefore vortex pinning and
remanence at solid walls can often be neglected in $^3$He-B and metastable states
with rapid superflow become possible. (iii) The most important difference lies in
the value of mutual friction. In $^4$He the dissipative part $\alpha$ is small and
influences vortex motion relatively little. Only within a narrow temperature
interval $\Delta T /T_{\lambda} \sim 2\cdot 10^{-3}$ from $T_{\lambda}$ one finds $q
> 1$ and expects different behavior. No conclusive measurements exist from so close
to $T_{\lambda}$.

In the helium superfluids, the transition at $q_{\rm c} \approx 1$ lies very close
to $T_{\lambda}$ in $^4$He, in the middle of the experimentally accessible
temperature range in $^3$He-B, and below that at extremely low temperatures in
$^3$He-A. This prediction is consistent with present experimental information. Here
we found that turbulence is unstable even at high $Re_{\rm s}$, if $q>q_{\rm c}$.
Perhaps in other multi-component hydrodynamic systems, superfluid or viscous, the
stability of turbulence might also be governed by an intrinsic parameter, in
addition to the Reynolds number.


This work was supported in part by the following  EU and ESF programmes: EU-IHP
ULTI-3, ESF-COSLAB, ESF-VORTEX. NK and GV are grateful to the Russian Foundation for
Basic Research; LS thanks for the grant GA\v{C}R (202/02/0251). We thank W.F. Vinen
for valuable discussion.

\end{document}